\documentclass[10pt, journal]{IEEEtran}

\usepackage[compress]{cite}
\usepackage{soul}
\usepackage{array}
\usepackage{amsmath}
\usepackage{amssymb,amsfonts}
\usepackage{textcomp}
\usepackage{stfloats}
\usepackage{cleveref}
\usepackage{lettrine}
\usepackage{multirow}
\usepackage{lipsum}
\usepackage{tikz}

\usepackage{algorithm}
\usepackage{algorithmic}

\usepackage{enumitem}
\usepackage{caption}
\usepackage{graphicx}
\begin{document}
\title{Deep Reinforcement Learning-Based Optimization of Second-Life Battery Utilization in Electric Vehicles Charging Stations}

\author{Rouzbeh Haghighi,~\IEEEmembership{Graduate Student Member,~IEEE,}
        Ali Hassan,~\IEEEmembership{Graduate Student Member,~IEEE,}
        Van-Hai Bui,~\IEEEmembership{Senior Member,~IEEE,}
        Akhtar Hussain,~\IEEEmembership{Member,~IEEE,}
        Wencong Su*,~\IEEEmembership{Senior Member,~IEEE,}
        \thanks{R. Haghighi, A. Hassan, V.-H. Bui, and W. Su are with the Department of Electrical and Computer Engineering, University of Michigan – Dearborn, Dearborn, MI 48128, USA.}
        \thanks{A. Hussain is with Laval University, Quebec City, Canada.}
        \thanks{* Corresponding author: Wencong Su (e-mail: wencong@umich.edu).}
        }
     
\markboth{}
{Author's Last Name \MakeLowercase{\textit{et al.}}: Title of Your Paper}

\maketitle
\begin{abstract}
The rapid rise in electric vehicle (EV) adoption presents significant challenges in managing the vast number of retired EV batteries. Research indicates that second-life batteries (SLBs) from EVs typically retain considerable residual capacity, offering extended utility. These batteries can be effectively repurposed for use in EV charging stations (EVCSs), providing a cost-effective alternative to new batteries and reducing overall planning costs. Integrating battery energy storage systems (BESS) with SLBs into EVCSs is a promising strategy to alleviate system overload. However, efficient operation of EVCSs with integrated BESS is hindered by uncertainties such as fluctuating EV arrival and departure times and variable power prices from the grid. This paper presents a deep reinforcement learning (DRL)-based planning framework for EV charging stations with BESS, leveraging SLBs. We employ the advanced soft actor-critic (SAC) approach, training the model on a year’s worth of data to account for seasonal variations, including weekdays and holidays. A tailored reward function enables effective offline training, allowing real-time optimization of EVCSs operations under uncertainty.
\end{abstract}

\vspace{-3mm}
\section*{Nomenclature}
{\small
\setlength{\tabcolsep}{2pt}
\renewcommand{\arraystretch}{1.3}
\begin{center}
\centering
\begin{tabular*}{\columnwidth}{l @{\extracolsep{\fill}} l}
\textbf{\textit{Indices:}} &  \\
\(t/d/k/sc\) & Set for Time, day, EV and scenario.  \\
\textbf{\textit{Constants:}} &  \\
\(C_{BESS}^{sc}\) & Capacity of BESS under scenario \(sc\) (Kwh). \\
\(\eta_{Ch/Dis}\) & Charging/discharging efficiency of the BESS. \\
\(SOC_{Max/min}\) & Maximum/minimum SOC of the BESS. \\
\multirow{2}{*}{\(N_{cycle}^{sc}\)} & Maximum remaining charge cycles of BESS \\
& in scenario \(sc\). \\ 
\(\psi_{LFP}\) & Cost of LFP (Lithium Iron Phosphate) battery. \\
\(\alpha\) & Degradation factor affecting BESS capital cost. \\
\(\mu\ / \sigma\) & Mean value / Standard deviation of the data. \\
\( N^{EV} \) & Total number of EVs \\
\( \mu_{com/pv} \) & Mean daily mileage for commercial/private EVs \\
\multirow{2}{*}{\( \sigma_{com/pv} \)} & Standard deviation of daily mileage for \\ 
 & commercial/private EVs. \\
\textbf{\textit{Variables:}} &  \\
\(P_t^{G}\)  & Power purchased from the utility grid at time \(t\). \\
\(P_t^{Ch/Dis}\)  & Power charged/discharged from the BESS at time \(t\). \\
\end{tabular*}
\end{center}}

{\small
\setlength{\tabcolsep}{2pt}
\renewcommand{\arraystretch}{1.3}
\begin{center}
\centering
\begin{tabular*}{\columnwidth}{l @{\extracolsep{\fill}} l}
\multirow{2}{*}{\(P_t^{EV_{Load}}\)} & Total load demand from all EVs in the parking lot \\
 & at time \(t\). \\
\(P_{t,k}^{EV}\) & Load demand of individual EV \(k\) at time \(t\). \\
\(\gamma_{Deg}\) & Degradation cost incurred during BESS operation. \\ 
\({Cap}_{BESS}^{sc}\) & Capital cost of BESS under scenario \(sc\). \\ 
\(x_{d}\) & Daily mileage of EVs on day \(d\). \\
\(f\) & Probability distribution function. \\
\(E_{d,k}^{need}\) & Total energy consumption of EV \(k\) on day \(d\). \\
\(\lambda^{EV}\) & Energy consumption rate of EVs. \\
\(\pi^G_t\) & Price of electricity from the utility grid at time \(t\). \\
\(R^{Total}\) & Total reward function in the SAC algorithm. \\
\(r_t^{PB}\) & Reward for enforcing the power balance constraint. \\
\multirow{2}{*}{\(r_t^{Ch/Dis}\)} & Reward for capturing the profit or cost of charging \\
 & /discharging the BESS. \\
\(r_t^{peak}\) & Penalty applied during peak time. \\
\(C_t^{Deg}\) & Penalty term associated with BESS degradation. \\
\(E_t^{cycle}\) & Penalty term associated the battery's end-of-life impact. \\
\end{tabular*}
\end{center}}

\section{Introduction}
In recent years, electric vehicles (EVs) have garnered significant attention for their potential to reduce carbon emissions, increasingly replacing traditional internal combustion vehicles. This shift is largely driven by escalating environmental concerns, including urban air pollution and climate change \cite{li2022comprehensive}. The growing market penetration of EVs has introduced new challenges regarding the management of millions of retired EV batteries, which may still retain substantial value. Currently, batteries with a state-of-health (SOH) between 70\% and 80\% are considered near the end of their useful life for EV applications, as they no longer meet the performance demands for driving \cite{hossain2019comprehensive}. These retired EV batteries, known as second-life batteries (SLBs), have certain disadvantages compared to new batteries, including lower maximum capacity, reduced charging/discharging efficiency, and a shorter remaining cycle life \cite{yang2021flexible}. The potential of SLBs within the power system, particularly in terms of economic impact and grid integration alongside SLB degradation, has been thoroughly investigated in \cite{hassan2023second}. Building on this, the market value of SLBs depends on various factors such as geography, SOH, and repurposing costs. An economic framework for estimating the market value of nickel cobalt oxide and lithium iron phosphate (LFP) SLBs in the U.S. and China through 2030 is presented in \cite{bach2024fair}.

The use of SLBs as battery energy storage systems (BESS) in EV charging stations (EVCSs) offers a promising approach to harness the whole potential of decommissioned batteries. In \cite{lin2023planning}, an innovative optimization framework for EVCSs is proposed, which considers the optimal integration of photovoltaic systems, wind power, BESS, and the use of SLBs. BESS generally face significant challenges in adapting to dynamic environmental changes, which limits their ability to identify optimal strategies \cite{haghighi2023cloud}. For effective operation, BESS must mitigate daily peak demand and prevent power system overload \cite{karimianfard2022economic}. However, their performance within EVCSs is influenced by multiple variables, including EV arrival and departure schedules, varying energy demands, and pricing from the upstream grid.
\vspace{-1mm}

Energy management optimization can be approached through various methods; however, machine learning techniques generally yield more efficient solutions than traditional algorithms. Reinforcement learning (RL) algorithms, in particular, have shown great potential, as they can dynamically design action strategies to maximize expected future rewards without prior knowledge of the environment. Unlike stochastic optimization, which produces probabilistic outcomes, RL’s model-free approach enables it to excel in managing systems with unknown dynamics and high levels of uncertainty, as demonstrated in numerous studies  \cite{lee2021dynamic}. In \cite{hu2024deep}, a hybrid energy system utilizing SLBs as energy storage is developed, along with a two-stage deep reinforcement learning (DRL) optimization approach to determine the best operational strategy in a dynamic setting. 
\vspace{-1mm}

The complexity of EVCS models increases with the number of random variables, making DRL a suitable approach for energy management in EVCS integrated with BESS. DRL effectively addresses the dynamic nature of the problem and the various uncertainties associated with EV demand prediction. Additionally, DRL models are resilient to minor perturbations, making them well-suited for handling modest uncertainties beyond the training data. As a result, the trained DRL model can be applied for real-time control of EVCS with BESS, reducing costs and mitigating the impact on the grid amidst uncertainties in EV arrival and departure times \cite{hussain2022deep, bui2019double}.
\vspace{-1mm}

Existing research has primarily focused on distinct aspects of cost reduction in EVCS or the operation of BESS, with some studies exploring the use of SLBs in power system applications. However, to address the challenges of uncertainty, high costs, and degradation, this paper proposes a framework utilizing SLBs as BESS in EVCS, aimed at reducing installation costs while accounting for degradation effects in BESS. This work introduces a DRL framework using the soft actor-critic (SAC) approach, selected for its rapid convergence, consistent performance, and ability to avoid local optima. This framework addresses a critical gap by optimizing both the operational and planning aspects of EVCS integrated with SLBs. The main contributions of this work are as follows:

\vspace{-2mm}
\begin{itemize}[leftmargin=3mm]
\item A SAC-based optimization framework for cost-efficient EVCS operation with BESS integration.
\item A innovative planning framework optimizing SLB configuration as BESS in EVCS to lower capital costs and environmental impact.
\item Incorporation of key degradation factors in BESS into the optimization problem.
\end{itemize}

\vspace{-4mm}
\section{System Architecture}
\vspace{-1mm}
\subsection{System Configuration}
Most EV owners prefer to charge their vehicles at home, leading to peak EV demand aligning with residential nighttime demand. This demand intensifies in areas with clustered EVs connected to a single charging station, potentially straining distribution transformers. To mitigate this, BESS can be used to reduce peak load on EV charging stations, helping to prevent transformer and power line overloading.

Fig. \ref{Fig_system} provides an overview of the system configuration in this study, which consists of three main areas: 1) utility grid, 2) EVCS, and 3) parking lots. As illustrated, various power flow combinations are possible for charging EVs in the parking lot. The EVCS can charge its BESS from the grid, supply power directly from the grid to the EVs, or discharge the BESS to meet EV demand. Eq. (\ref{eq:powerB}) ensures power balance at each time interval in the EVCS, where the power purchased from the grid and discharged from the BESS must equal the sum of the power demand for charging the BESS and supplying the total EV load. The total EV load is calculated using Eq. (\ref{eq:EV_Load}), as the sum of individual EV loads. The integration of BESS allows the EVCS to charge the battery during off-peak hours and discharge it during peak demand periods. However, this strategy requires careful consideration of battery degradation costs, including the impact of charging and discharging on SOH, which is discussed in Section II.B.

\vspace{-3mm}
{\small
\begin{equation}
\label{eq:powerB}
    P_t^{G} + P_t^{Dis} = P_t^{EV_{Load}} + P_t^{Ch} \quad \quad \quad \quad \quad \quad \quad \quad \quad \quad \forall t 
\end{equation}
\begin{equation}
\label{eq:EV_Load}
    P_t^{EV_{Load}} = \sum_{k=1}^{N^{EV}} P_{t,k}^{EV} \quad \quad \quad \quad \forall t , \forall k \in \{1,2,3, ..., N^{EV}\}
\end{equation}}

\vspace{-4mm}
{\setlength{\belowcaptionskip}{-5mm}
\begin{figure}[!b]
\centering
\includegraphics[width=0.9\linewidth]{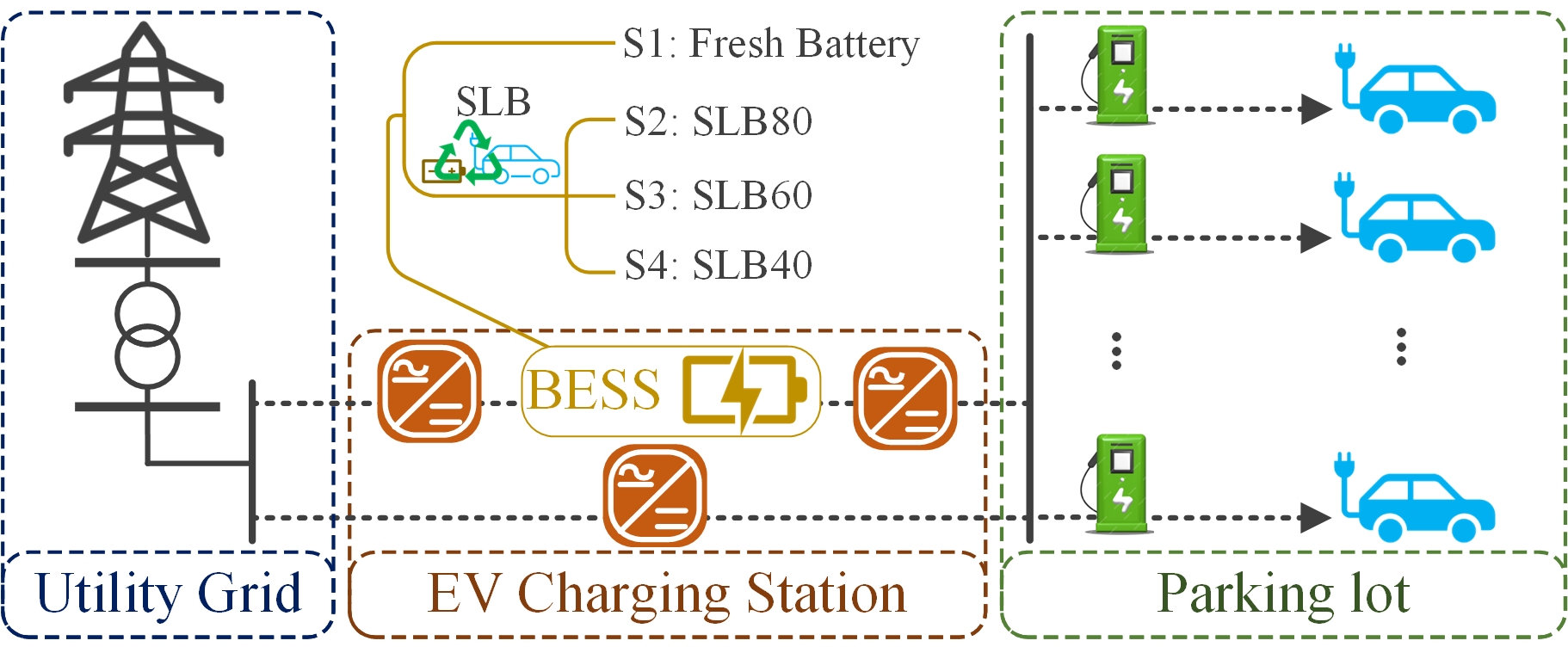}
\caption{Configuration of the EVCS with BESS.}
\label{Fig_system}
\end{figure}}

\vspace{-2mm}
\subsection{Battery Energy Storage Model}
BESS modeling is essential for realistic planning and operation of EVCS. The energy stored in the BESS is determined by Eq. (\ref{eq:SOC}), representing the sum of the initial energy and the cumulative power charged or discharged up to interval \(t\). Additionally, the SOC must remain within a specified range to prevent damage to the BESS, as shown in Eq.  (\ref{eq:SOCMm}). Eqs. (\ref{eq:SOCCh})-(\ref{eq:SOCDis}) impose limits on the BESS charging and discharging rates, with efficiency coefficients applied to ensure SOC consistency and reflect actual energy transfers.

\vspace{-4mm}
{\small
\begin{equation}
    \label{eq:SOC}
    SOC_{t+1} = SOC_t + P_t^{Ch} - P_t^{Dis} \quad \quad \quad \quad \quad  \quad \quad \quad \quad \quad \forall t
\end{equation}
\begin{equation}
    \label{eq:SOCMm}
    SOC_{min} \leq SOC_t \leq SOC_{Max} \quad \quad \quad \quad \quad \quad \quad \quad \quad \quad \quad \forall t
\end{equation}
\begin{equation}
    \label{eq:SOCCh}
    0 \leq P_t^{Ch} \leq C_{BESS}^{sc} \cdot \left(\frac{1 - SOC_t}{\eta_{Ch}} \right) \quad \quad \quad \quad \quad \quad \quad \quad \quad \forall t
\end{equation}
\vspace{-3mm}
\begin{equation}
    \label{eq:SOCDis}
    0 \leq  P_t^{Dis} \leq C_{BESS} \cdot SOC_t \cdot {\eta_{Dis}} \quad \quad \quad \quad  \quad \quad \quad \quad \quad \quad \forall t
\end{equation}}

An optimal energy management strategy should incorporate the degradation costs of BESS into the scheduling and operational processes \cite{amini2023predictive}. This study considers the integration of various BESS types, including fresh BESS and SLBs, into the EVCS. The degradation costs differ between these batteries and are calculated using Eq. (\ref{eq:Deg}), based on the total number of cycles and the capital cost of each BESS. \cite{bach2024fair} shows that in the U.S. market, LFP SLB prices range from 40\% to 75\% of the fresh battery price, depending on SOH and projected degradation paths. Since this study considers LFP batteries, Eq. (\ref{eq:cap}) calculates the capital cost of each BESS type based on a degradation factor (\(\alpha\)). Fig. 1 in \cite{Github_repository} shows the normalized usable capacity and degradation capacity of different BESS types examined in this study. It is assumed that each battery operates from its initial capacity to End of Life capacity, which is considered 20\% of its nominal capacity. The total degradation cost per cycle is given by total capacity degradation by total number of cycles. It should be noted that, in reality, battery degradation follows a nonlinear pattern. However, for simplicity, it is approximated as a linear process in this study.

\vspace{-3mm}
{\small
\begin{equation}
    \label{eq:Deg}
    \gamma_{Deg} = \frac{Cap_{BESS}^{sc}}{num_{cycle}}
\end{equation}
\begin{equation}
    \label{eq:cap}
    Cap_{BESS}^{sc} = \alpha \times C_{BESS}^{sc} \times \psi_{LFP}
\end{equation}}

\vspace{-5mm}
\subsection{EV Load Modeling}
To achieve optimal operation of the EVCS, the total EV load in the parking lot at each time interval must be determined. This load varies due to factors such as daily mileage, energy consumption per kilometer, and the State of Charge (SOC). Daily mileage also depends on vehicle use—for instance, commercial EVs generally travel more on weekdays than private vehicles. As a result, daily EV mileage is modeled as a random variable following a lognormal distribution (\(f(x_{d} | \mu, \sigma)\)), represented Eq. (\ref{eq:Norm}), where \(x\) denotes daily mileage. 

\vspace{-3mm}
{\small
\begin{equation}
\label{eq:Norm}
f(x_{d} | \mu, \sigma) = \frac{1}{x_{d} \cdot \sigma \cdot \sqrt{2\pi}} e^{-\frac{(\ln x_{d} - \mu)^2}{2\sigma^2}}, \quad x_{d} > 0
\end{equation}}

Fig. 2 in \cite{Github_repository} illustrates the probability density functions of daily mileage for private and commercial EVs, modeled as lognormal distributions, as supported by several studies \cite{hussain2020optimal}. Commercial EV loads are distributed between 8 AM and 6 PM (10 hours), while private EV loads are concentrated between 6 PM and 10 PM (4 hours). This approach captures general usage trends; however, the model could be refined by introducing slight variations in start and end times within these windows to better simulate real-world charging behaviors. Additionally, if data on typical charging times were available, a probability-based approach could further enhance accuracy. The daily energy requirement for each EV is calculated as Eq. (\ref{eq:daily}). This model assumes that each EV charges in a single session to meet its daily mileage needs. To enhance accuracy, SOC levels can be factored into load calculations, scheduling charging sessions only when SOC falls below a certain threshold (e.g., 50\%), especially for private EVs that may not charge daily.

\vspace{-3mm}
{\small
\begin{equation}
\label{eq:daily}
    E_{d,k}^{need}= \frac{x_{d}}{\lambda^{EV}}
\end{equation}}

\vspace{-4mm}
\section{DRL-based Optimization Framework}
\vspace{-2mm}
The optimal operation of BESS in EVCS is a complex multi-period problem that requires consideration of past SOC data and future load demands. While stochastic optimization can address this, it is limited by its reliance on complete knowledge of stochastic processes and becomes computationally infeasible with an increasing number of EVs due to the numerous random variables involved. This study proposes a DRL-based approach that eliminates the need for system modeling or approximating uncertainties. Instead, it learns optimal strategies over time by interacting directly with the environment.
\vspace{-1mm}

Fig. \ref{Fig_SAC} illustrates the flowchart of the proposed SAC-based model for optimizing EVCS operations with different scenarios of BESS. The method aims to determine the optimal charging/discharging schedule by utilizing the SAC architecture: the critic evaluates the policy by updating the action-value function, which measures the net discounted reward, while the actor interacts with the environment by selecting actions based on the current state. The detailed implementation steps of the SAC algorithm are outlined in Algorithm~\ref{alg:sac}.
The SAC optimization framework is constructed based on the following definitions:

{\setlength{\belowcaptionskip}{-7mm}
\begin{figure}[!t]
\centering
\includegraphics[width=0.8\linewidth]{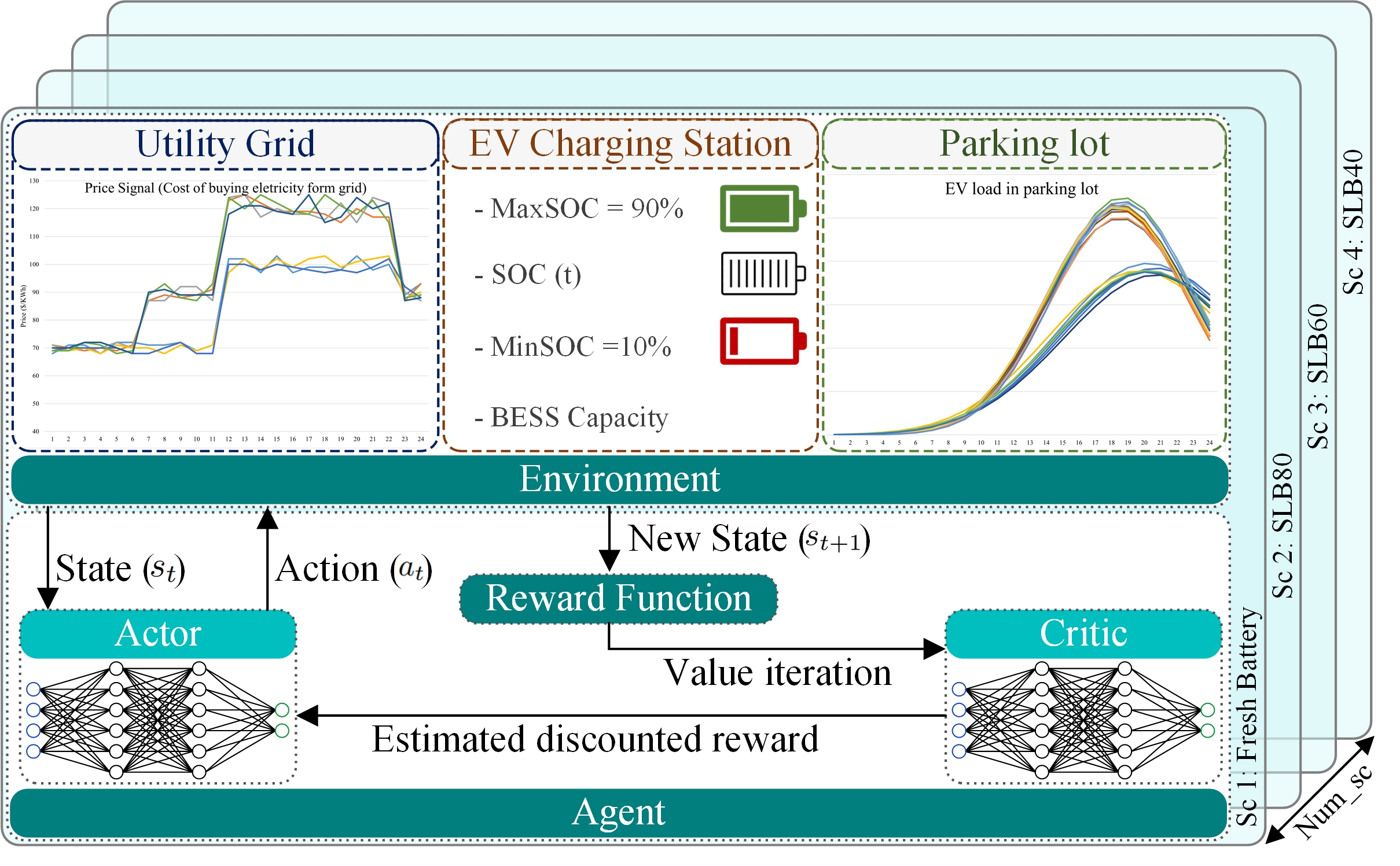}
\caption{Flowchart of the proposed SAC-based framework for EVCS operation.}
\label{Fig_SAC}
\end{figure}}

\begin{itemize}[leftmargin=3mm]
\vspace{-1mm}
    \item \textbf{State:} The environment state of the SAC is defined based on \( s_t \in S \), where the state at time \( t \), is represented by: \(s_t = [t,  SOC_t, \pi^G_t, P_t^{EV}] \). Here $t$ is the current time of the day; $SOC_t$ is the amount of stored energy in the BESS;  $\pi^G_t$ is the current electricity price; and finally, $ P_t^{EV}$ is the current energy consumption of EVs (i.e., the total demand from the EVs in the parking lot). The state function will be normalized using a min-max normalization technique.
    
    \item \textbf{Action:} The actions in the SAC framework are taken to modify the system state. The actions are represented as a vector \( a_t = [a_1, a_2, a_3] \), where \( a_1 \) represents the amount of energy to charge the BESS from the grid, \( a_2 \) is the amount of energy to discharge the BESS to charge the EVs, and \( a_3 \) is the amount of energy to charge the EVs directly from the grid. The action space is constrained by the maximum charging/discharging capacity of the BESS and the power converter's capacity for charging EVs directly from the grid.

    \item \textbf{Reward:} The reward function serves as the core of any RL algorithm, guiding the agent's actions to make optimal decisions based on load, pricing, and system efficiency. This study aims to minimize EVCS operational costs by efficiently managing BESS to reduce peak load during peak hours while accounting for battery degradation costs. Electricity prices reflect power system load, typically higher during peak periods and lower during off-peak times. Consequently, market prices are used to identify system peak load intervals. The reward function, incorporating these factors, is calculated using Eq. (\ref{eq:Total_r}).
\end{itemize}
\vspace{-2mm}
{
\setlength{\belowcaptionskip}{-25mm}
\begin{algorithm}[ht]
\caption{Soft Actor-Critic (SAC) for EVCS Scheduling}
\label{alg:sac}
\begin{algorithmic}[1]
\STATE \textbf{Initialize} environments $\text{Env}_{sc}$ for each BESS scenario and corresponding SAC agents
\STATE \textbf{Set} training parameters (e.g., batch size, learning rates)
\FOR{each scenario $sc$}
    \STATE \textbf{Initialize} total operational cost $C_{\text{operation}} \leftarrow 0$
    \STATE \textbf{Initialize} actor network $\pi_\theta$, critic networks $Q_{\phi_1}$, $Q_{\phi_2}$, value network $V_{\psi}$, and target value network $\hat{V}_{\psi'}$
    \STATE \textbf{Initialize} empty replay buffer ($\mathcal{D}$)
    \FOR{each episode}
        \STATE Reset environment and receive initial state $s_0$
        \STATE \textbf{Initialize} episode reward $R \leftarrow 0$
        \WHILE{not done}
            \STATE Select action $a_t$ according to policy $\pi_\theta(a_t|s_t)$
            \STATE Execute action $a_t$, observe reward $r_t$, next state $s_{t+1}$, and done signal $d_t$
            \STATE Store transition $(s_t, a_t, r_t, s_{t+1}, d_t)$ in replay buffer
            \STATE Sample a mini-batch of transitions from $\mathcal{D}$
            \STATE Compute target values and critic losses
            \STATE Update critic networks ($Q_{\phi_1}$, $Q_{\phi_2}$) using gradient descent
            \STATE Compute policy loss and update actor network $\pi_\theta$ using gradient ascent
            \STATE Compute value function loss and update value network $V_{\psi}$
            \STATE Update state $s_t \leftarrow s_{t+1}$
            \STATE Accumulate reward: $R \leftarrow R + r_t$
            \STATE Update policy network using gradient ascent to maximize expected return
        \ENDWHILE
        \STATE Accumulate total operational cost $C_{\text{operation}}$
        \STATE Save model weights periodically
    \ENDFOR
    \STATE Calculate total cost and net capital cost impact in $sc$
\ENDFOR
\end{algorithmic}
\end{algorithm}}

\vspace{-3mm}
The reward function in this study is designed to achieve optimal BESS operation under varying conditions. The power balance component (\( r_t^{PB} \)) minimizes the difference between the threshold load level (\( P^{Th} \)) and the net demand of the EVCS with BESS, as defined in Eq. (\ref{eq:r_pb}). To incentivize charging during low-price intervals, Eq. (\ref{eq:r_ch}) calculates a reward based on a comparison between the current interval's price (\( \pi^G_t \)) and a randomly selected price from the next \( i \) intervals (\( \pi^G_i \)). Similarly, for discharging, Eq. (\ref{eq:r_dis}) provides a reward for discharging during peak price intervals, with future price information incorporated in the same way as the charging reward. This approach ensures the agent develops an optimal policy to maximize profit by charging during off-peak periods and discharging during peak periods. To address peak load management, Eq. (\ref{eq:r_peak}) introduces a penalty for deviations from the threshold load level during peak load intervals. This penalty is zero during off-peak periods but doubles during peak load periods—once due to the baseline power balance factor and again due to the peak load penalty—ensuring the system stays close to the threshold and minimizes grid purchases during peak periods. Additionally, Eq. (\ref{eq:r_deg}) imposes a penalty for energy traded to account for BESS degradation over time. Eq. (\ref{eq:E_cycle}) calculates the total number of cycles for each BESS scenario, ensuring that the cycles remain within the specified limits for each case. Finally, the total reward function is normalized by the number of days (\( D \)) and intervals per day (\( \Gamma \)) to ensure consistent analysis across different training horizons. In this equation, T represents the total time for each episode. This comprehensive reward structure balances cost efficiency, peak load management, and battery degradation in BESS operations.

\vspace{-5mm}
{\small
\begin{equation}
\label{eq:Total_r}
R^{Total} = \sum_{t=1}^{T} \left( \frac{r_t^{PB} + r_t^{Ch} + r_t^{Dis} + r_t^{peak} - C_t^{Deg} + E_t^{cycle}}{D \cdot \Gamma} \right)
\end{equation}
\vspace{-3mm}
\begin{equation}
\label{eq:r_pb}
r_t^{PB} = - ((P_t^{EV} + P_t^{Ch} - P_t^{Dis}) - P^{Th}) ^2
\end{equation}
\vspace{-3mm}
\begin{equation}
\label{eq:r_ch}
r_t^{Ch} = P_t^{Ch} \cdot ( {rand}(\pi^G_i) - \pi^G_t ) \quad \quad \quad \quad \quad \quad \quad i \in [t, \Gamma]
\end{equation}
\begin{equation}
\label{eq:r_dis}
r_t^{Dis} = P_t^{Dis} \cdot ( \pi^G_t - {rand}(\pi^G_i) ) \quad \quad \quad \quad \quad \quad i \in [t, \Gamma]
\end{equation}
\begin{equation}
\label{eq:r_peak}
r_t^{peak} = 
\begin{cases} 
      r_t^{PB} & \quad \quad \quad \quad \text{if } t \in \left[ t^{pb}, t^{pe} \right] \\
      0 & \quad \quad \quad \quad \text{else}
\end{cases}
\end{equation}
\begin{equation}
\label{eq:r_deg}
C_t^{Deg} = \gamma_{Deg} \cdot ( P_t^{Ch} + P_t^{Dis} )
\end{equation}
\begin{equation}
\label{eq:E_cycle}
E_t^{cycle} = 
\begin{cases} 
      -1000 & \quad \quad \quad \quad \text{if } cycle \geq N^{sc}_{cycle} \\
      0 & \quad \quad \quad \quad \text{else}
\end{cases}
\end{equation}}

\vspace{-5mm}
\section{Results and Discussions}
This study considers two types of EVs: commercial and private. The EV load is modeled as described in Section II.C, with the corresponding parameters summarized in Table I in \cite{Github_repository}. The hyperparameters used for simulation are listed in Table II in \cite{Github_repository}. Real market energy prices are calculated using electricity rates from DTE Energy \cite{dte_pricing}. The capital cost of LFP batteries is assumed to be  $\psi_{LFP}=\$389/kWh$ for fresh batteries. For SLBs, the capital cost is reduced based on degradation levels, with decreasing factors of 75\%, 57\%, and 40\% applied for SLB80, SLB60, and SLB40, respectively. The total cycle counts for LFP batteries in this study are 15,000 for fresh batteries, 10,000 for SLB80, 7,500 for SLB60, and 5,000 for SLB40. The SAC, DDPG, and TD3 implementations were executed and evaluated on an Intel Core i9-12900K CPU, an NVIDIA RTX 3090 GPU (24GB VRAM), and 64GB RAM using Python in the Spyder environment.

The reward function of the SAC algorithm (Fig. \ref{Fig_reward}) demonstrates convergence across all scenarios, indicating the model's ability to consistently optimize the policy for each battery type. As shown in Fig. \ref{Fig_r4}, the action frequency indicates that in the Fresh Battery scenario, the agent primarily charges EVs directly from the grid to avoid high degradation costs. In contrast, for other scenarios, particularly SLB60, it strategically utilizes the BESS for charging/discharging to optimize costs and efficiency. The total cost breakdown (Fig. \ref{Fig_r3}), which includes operational costs, degradation, capital costs, and total cost highlights significant variations across the scenarios. Fresh batteries exhibit higher costs due to their capital expenses and degradation impact, while SLBs demonstrate reduced overall costs, showcasing the economic advantages of utilizing SLBs as BESS in EVCS operations.
\vspace{-4mm}

{\setlength{\belowcaptionskip}{-8mm}
\begin{figure}[ht]
\centering
\includegraphics[width=0.8\linewidth]{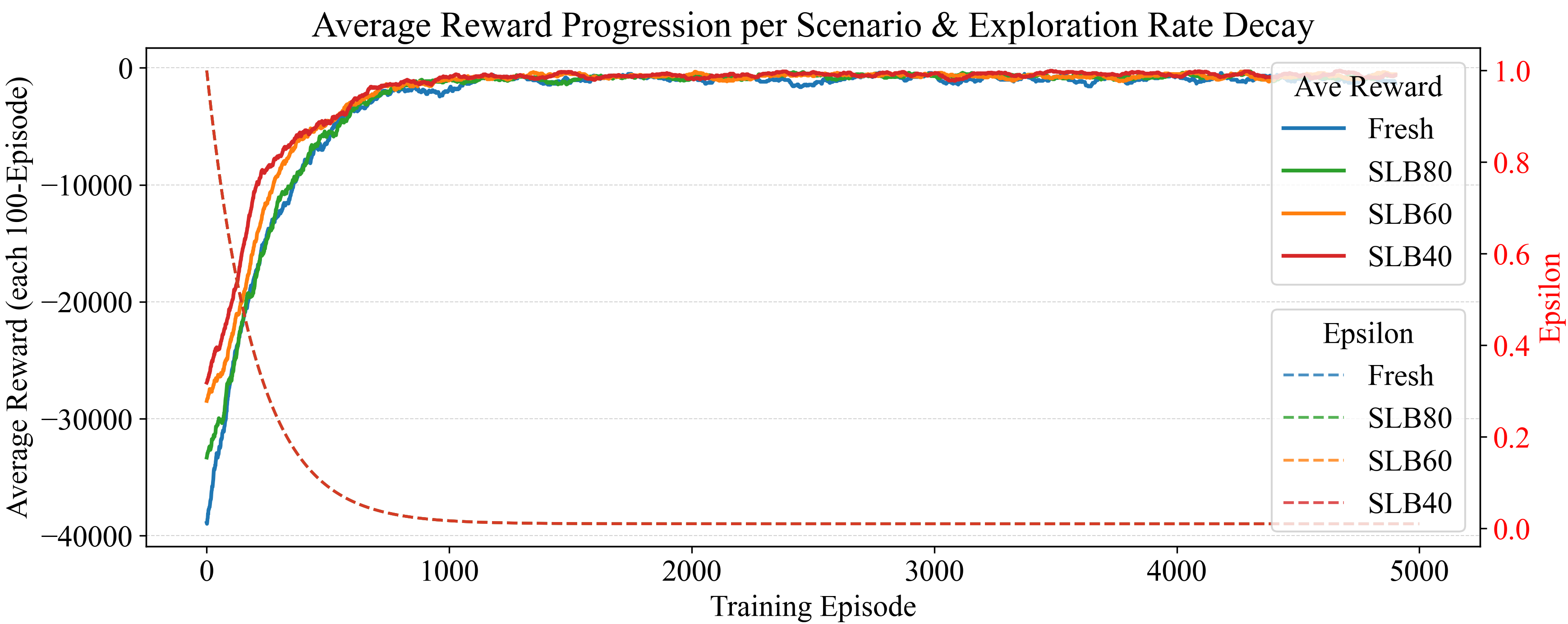}
\caption{Convergence analysis of SAC and epsilon decay during training.}
\label{Fig_reward}
\end{figure}}

{\setlength{\belowcaptionskip}{-8mm}
\begin{figure}[ht]
\centering
\includegraphics[width=0.8\linewidth]{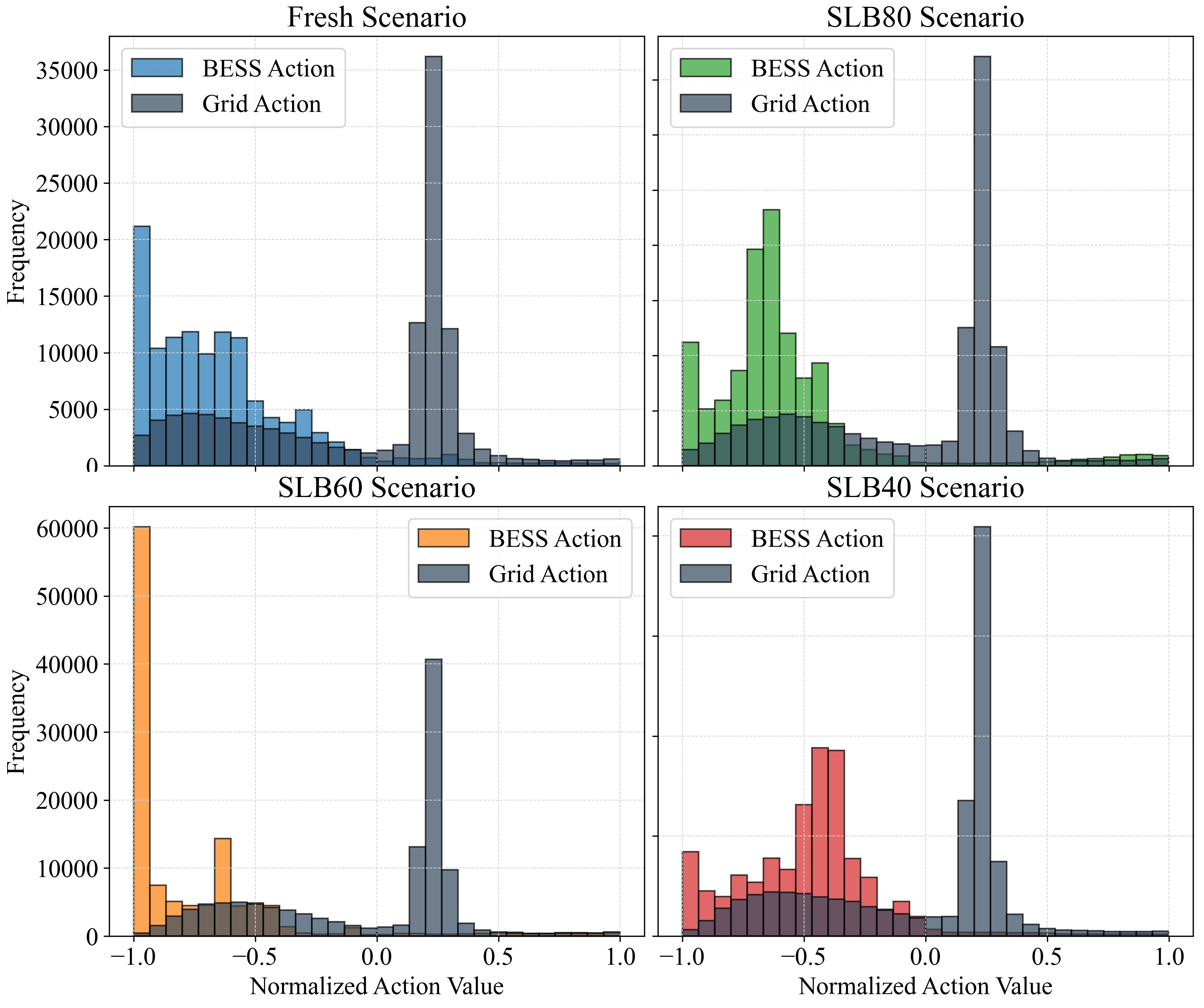}
\caption{Frequency of actions chosen by the SAC (BESS: [-1,1], Grid: [0,1]).}
\label{Fig_r4}
\end{figure}}

{\setlength{\belowcaptionskip}{-3mm}
\begin{figure}[ht]
\centering
\includegraphics[width=0.9\linewidth]{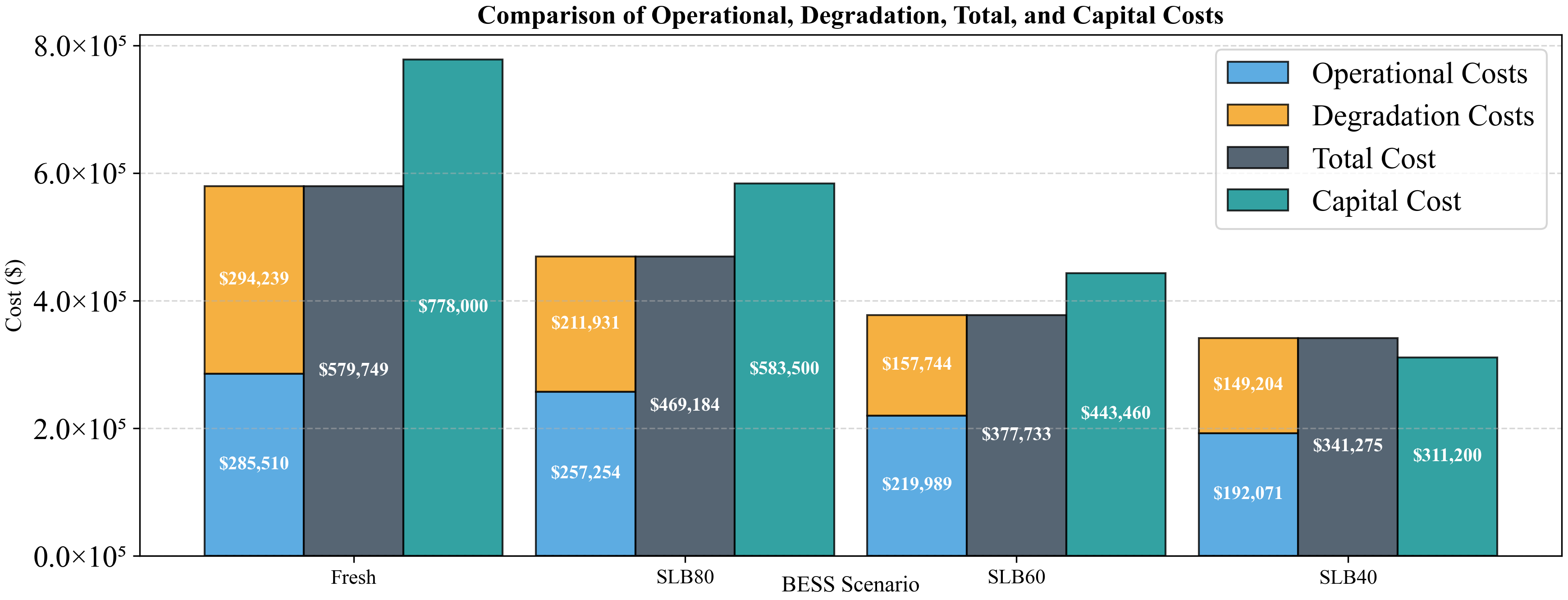}
\caption{Total cost breakdown across battery scenarios.}
\label{Fig_r3}
\end{figure}}

To evaluate SAC's effectiveness as an optimization method, its performance in the SLB80 scenario is compared with the state-of-the-art DRL method, Twin Delayed DDPG (TD3), as shown in Fig. \ref{Fig_r4}. SAC demonstrates superior performance with higher rewards during training, lower operational costs, and better policy determination for BESS operation. Additionally, SOC violation analysis (Fig. \ref{Fig_r4}) confirms SAC’s advantage, as it exhibits fewer violations, ensuring more reliable performance.

\vspace{-8mm}
\section{Conclusions}
This study explores the utilization of SLBs in EVCS to highlight their advantages in terms of cost-effectiveness, economic feasibility, and environmental impact. By analyzing various scenarios involving SLBs and fresh batteries as BESS, with differing capital and degradation costs, we demonstrated the significant benefits of integrating SLBs. To ensure accurate and efficient planning, we employed the SAC algorithm, a DRL approach, to determine the optimal operational strategy for EVCS. A comprehensive reward function was developed to account for the problem's constraints and objectives. Simulation results validated the effectiveness of the proposed framework, showcasing its superiority in optimizing EVCS operations and demonstrating the value of SLBs as a viable BESS solution.

{\setlength{\belowcaptionskip}{-9mm}
\begin{figure}[ht]
\centering
\includegraphics[width=0.7\linewidth]{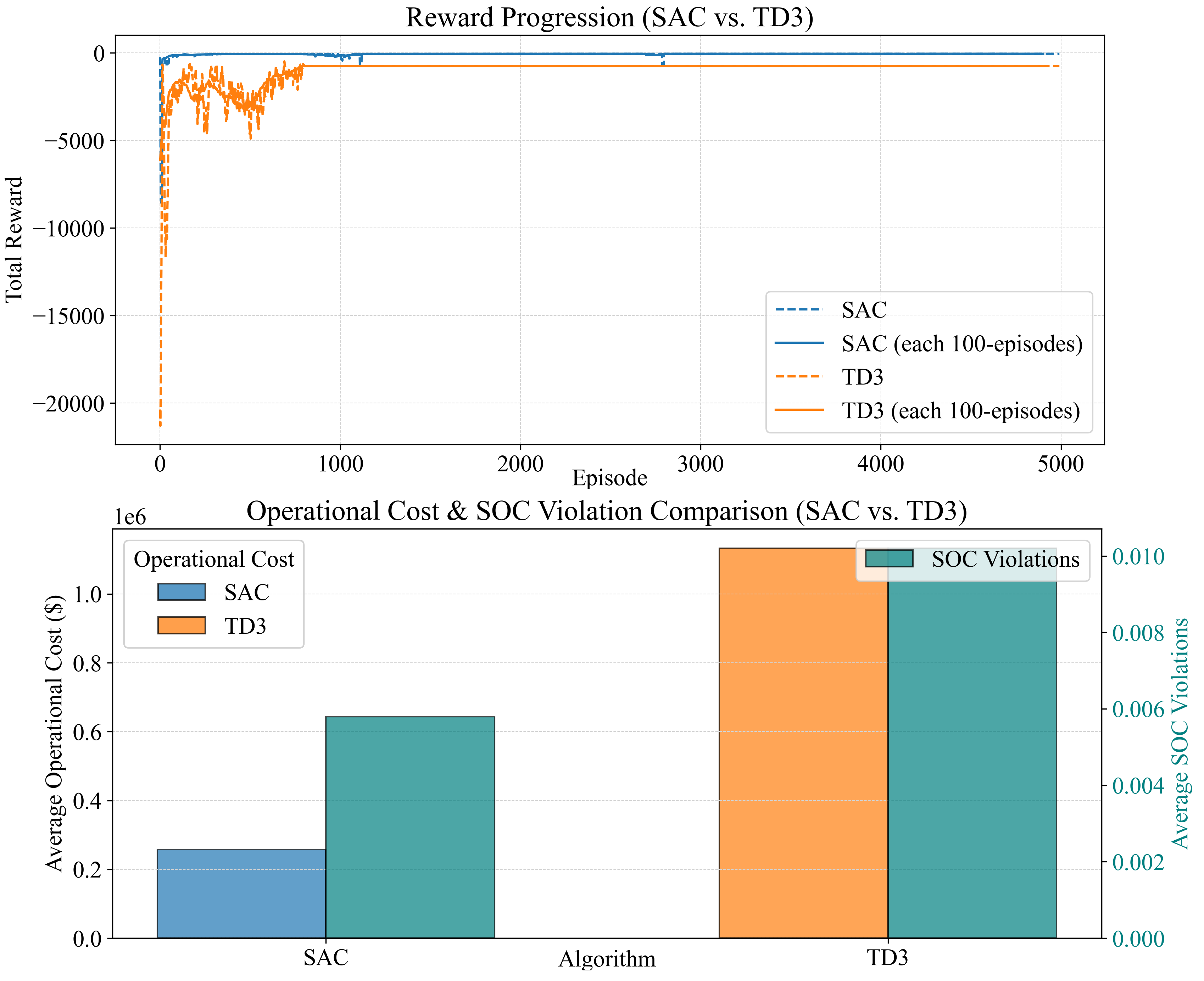}
\vspace{-2mm}
\caption{Reward progression, cost comparison and SOC violation results of SAC and TD3.}
\label{Fig_r4}
\end{figure}}

\vspace{-4mm}
\footnotesize
\bibliographystyle{IEEEtran}
\bibliography{Ref}
\end{document}